\documentstyle[12pt,epsfig]{article}
\begin{document}

\title{\Large 
{\normalsize     
IC/99/36\hfill\mbox{}\\
April 1999\hfill\mbox{}\\}
\vspace{.8cm}
\bf NEUTRINO SPECTRUM,\\ 
OSCILLATION SCENARIOS AND\\ 
NEUTRINOLESS DOUBLE BETA DECAY}
%
%\vspace{2cm} 
\author{Francesco Vissani\\
{\it Deutsches Elektronen Synchrotron, DESY,}\\
{\it Notkestra\ss{}e 85, D-22603 Hamburg, Germany;}\\
{\it International Centre for Theoretical Physics, ICTP}\\
{\it Strada Costiera 11, 34100 Trieste, Italy}}
\date{}
\maketitle

\begin{abstract}
We introduce the representation 
on one unitarity triangle of the constraints resulting 
(1) from the interpretation of solar and atmospheric 
neutrino data in terms of oscillations,
and (2) from the search for neutrinoless 
double beta decay. We show its use  
for the study of a nearly degenerate 
neutrino spectrum. 
The representation shows clearly the particular
cases when the neutrinoless double beta decay rate 
can (or cannot) be small, that is: 
when the connection of the decay rate 
with the neutrino spectrum is less (or more) direct. 
These cases turn out to depend crucially on the 
scenario of oscillation (MSW solutions, vacuum
oscillations, averaged oscillations), 
and in particular on the size of the mixing between the
electron neutrino and the neutrino state giving rise to 
atmospheric neutrino oscillations.
\end{abstract}

Atmospheric neutrino data can be interpreted in terms of 
dominant $\nu_\mu-\nu_\tau$ oscillation channel; 
a sub-dominant channel $\nu_\mu-\nu_{\rm e}$ 
is not excluded. 
In terms of the mixing elements, 
we can summarize these informations by
$|U_{\rm \tau 3}^2| \sim |U_{\rm \mu 3}^2|
\gg |U_{\rm e3}^2|,$
assuming that the 
heaviest mass $m_3$ is the one responsible 
of atmospheric neutrino oscillations. 

Several possibilities are opened for interpretation of
the solar neutrino data, depending on the 
frequencies of oscillation and mixings.

However,  there is still 
quite a limited knowledge 
on the {\em neutrino mass spectrum itself.}
The search for neutrinoless double beta 
($0\nu 2\beta$) decay can shed light on this issue.
The bound obtained on the parameter
${\cal M}_{{\rm ee}}=| \sum_i U_{{\rm e}i}^2\, m_i |$ 
($m_3\ge m_2\ge m_1$)  is sensibly smaller than the mass scales 
probed by present studies of $\beta$-decay, or those 
inferred in cosmology.

The largest theoretical value 
is  taken for a nearly degenerate spectrum 
$m_1\approx m_2 \approx m_3$ \cite{[1],[2]};
the value is\footnote{Notice however that
a value of ${\cal M}_{{\rm ee}}\sim 
(\Delta m^2_{atm})^{1/2}$ would 
be already quite large on its own.} 
${\cal M}_{{\rm ee}}\approx m_1,$  with 
good approximation 
if $m_1\gg (\Delta m^2_{atm})^{1/2}.$ 
The corresponding minimum value, 
under an arbitrary variation of 
the unknown phases, is conveniently 
represented in an unitarity triangle (fig 1).
{}From the figure it is visible that,
to interpret properly the results of 
$0\nu 2\beta$ decay studies 
(and possibly, to exclude the inner region 
in $1^{st}$ plot, the one where 
${\cal M}_{{\rm ee}}\ll m_1$
is {\em possible}) 
we need precise informations 
on the mixing elements.
This requires 
distinguishing among 
oscillation scenarios. 
The plots illustrate also
the importance of quantifying 
the size of $|U_{\rm e3}^2|$ \cite{[1],[2]}, \cite{[3]}.
In the case in which  the state responsible of neutrino
oscillations is the lightest one--not the heaviest--, 
very similar considerations apply. Graphically,
this case corresponds to a $120^\circ$  rotation 
of the last three plots of fig 1
(the sub-dominant mixing being $|U_{\rm e1}^2|$).
\vskip.3cm
\noindent
{\Large\bf Acknowledgements}\\[1ex]{} 
I thank for useful discussions 
R Barbieri, C Giunti, M Maris
and A Yu Smirnov.

\newpage
\begin{figure}
\begin{center}
\epsfig{file=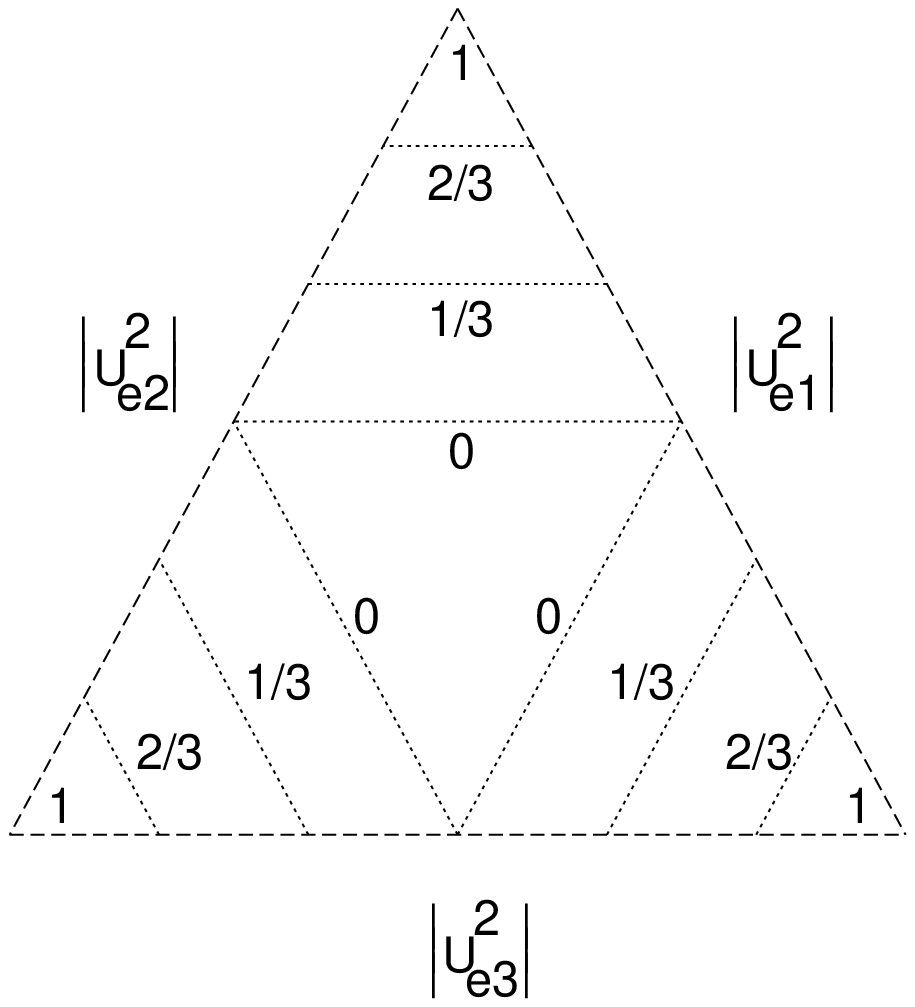, angle=270,
width=7.7cm}\epsfig{file=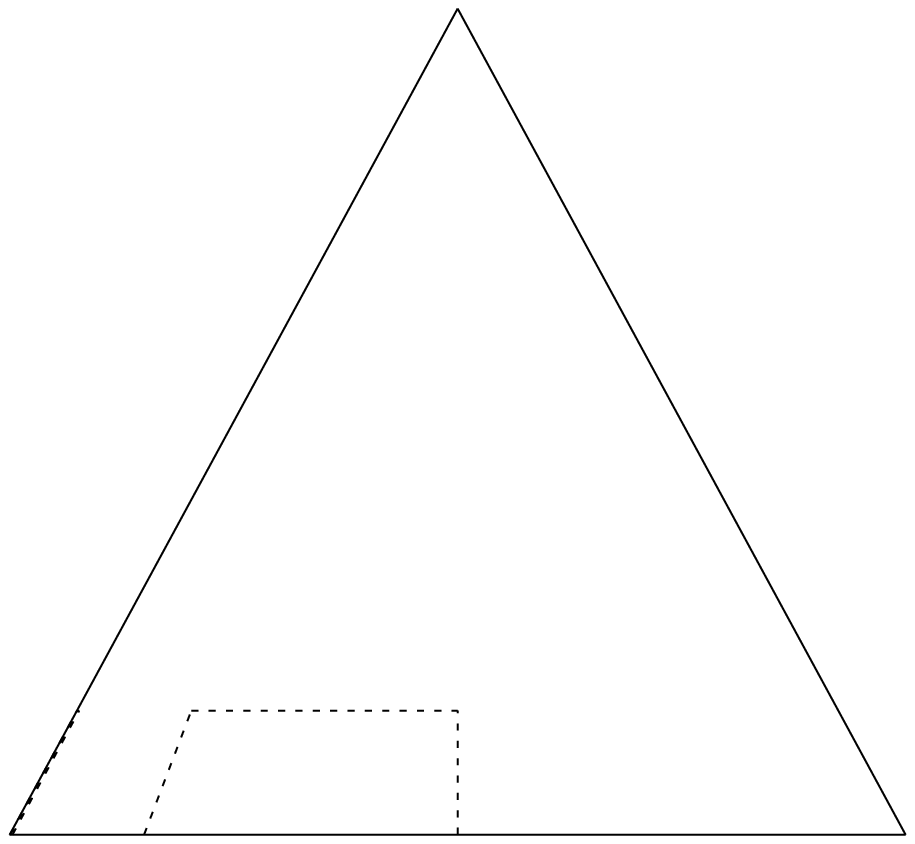, angle=270,width=7.7cm}
\epsfig{file=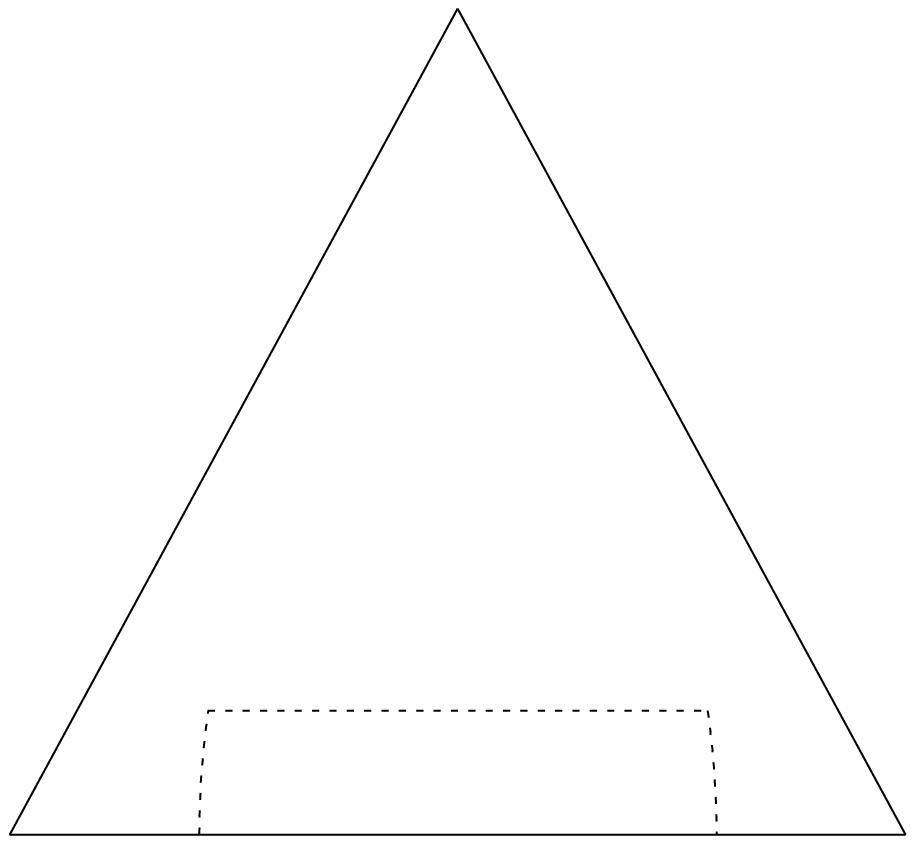, angle=270,width=7.7cm}\epsfig{file=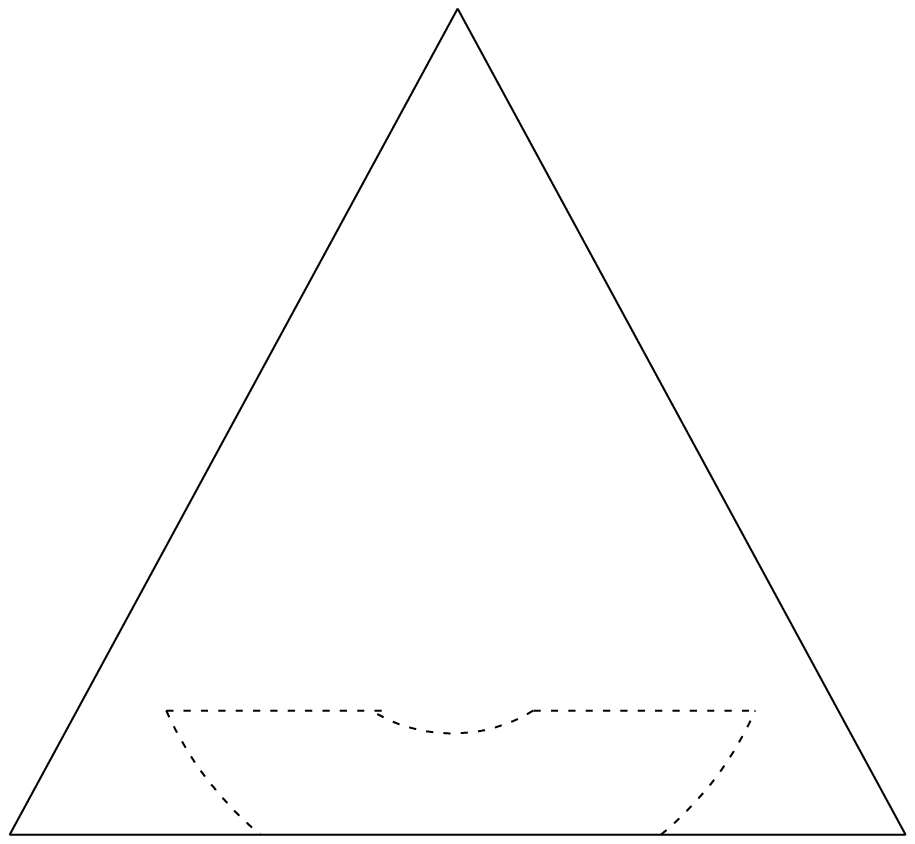, 
angle=270,width=7.7cm}
\end{center}
\caption{The minimum value of ${\cal M}_{{\rm ee}}/m_1,$
${\rm max}\{ 2 |U_{{\rm e}i}^2|-1,\  0 \},$
in function of the mixing angles (see \cite{[2]}). 
{}From up to down, left to right:
$1^{st}$ plot, 
{\em legenda} and numerical values
of reference;
$2^{nd}$, indicative region for
MSW enhanced transitions;
$3^{rd},$ for vacuum oscillations; $4^{th},$ for 
averaged oscillations. 
We assumed $|U_{\rm e3}^2|<0.15$ \cite{[3]}.}
\end{figure}
\end{document}